\documentclass[twocolumn,showpacs,prb]{revtex4}
\usepackage{graphics,german,amsmath,amssymb}

\usepackage{graphicx}
\usepackage{color}
\usepackage{bm}

\newcommand{\mvec}[1]{{\bf #1}}
\newcommand{\mmatrix}[1]{\underline{\underline{#1}}}

\definecolor{olive}{rgb}{0.0,0.5,0.0}
\definecolor{darkred}{rgb}{0.5,0.0,0.0}

\begin{document}

\title{Antiferromagnetic Dimers of Ni(II) in the $S$~=~1 Spin-Ladder  Na$_2$Ni$_2$(C$_2$O$_4$)$_3$(H$_2$O)$_2$}
\author{C. Mennerich$^{1}$}
\author{H.-H. Klauss$^{1}$}
\author{M. Broekelmann$^{1,2}$}
\author{F.J.~Litterst$^{1}$}
\author{C. Golze$^{1,2}$}
\author{R. Klingeler$^{2,4}$}
\author{V. Kataev$^{2}$}
\author{B. B"uchner$^{2}$}
\author{S.-N. Grossjohann$^{3}$}
\author{W. Brenig$^{3}$}
\author{M. Goiran$^{4}$}
\author{H. Rakoto$^{4}$} 
\author{J.-M. Broto$^{4}$}
\author{O. Kataeva$^{5}$}
\author{D.J. Price$^{6}$}

\affiliation{$^{1}$Institut f\"ur Physik der Kondensierten Materie, TU Braunschweig, Mendelssohnstr.3, D-38106 Braunschweig, Germany}
\affiliation{$^{2}$Leibniz-Institute for Solid State and Materials Research IFW Dresden, P.O. Box 270116, D-01171 Dresden, Germany}
\affiliation{$^{3}$ Institut f\"ur Theoretische Physik, TU Braunschweig, Mendelssohnstr.3, D-38106 Braunschweig, Germany}
\affiliation{$^{4}$ Laboratoire National des Champs Magn\'{e}tiques Puls\'{e}s, 31432 Toulouse Cedex 04, France}
\affiliation{$^{5}$ Arbuzov Institute of Organic and Physical Chemistry, RAS, 420088 Kazan, Russia}
\affiliation{$^{6}$ WestCHEM,Department of Chemistry, University of Glasgow, Glasgow, G12 8QQ, UK}
\date{13th January 2006}

\begin{abstract}

We report the synthesis, crystal structure and magnetic properties of the $S$=1 2-leg spin-ladder compound Na$_2$Ni$_2$(C$_2$O$_4$)$_3$(H$_2$O)$_2$. The magnetic properties were examined by magnetic susceptibility and pulsed high field magnetization measurements. The magnetic excitations have been measured in high field high frequency ESR. Although the Ni(II) ions form structurally a 2-leg ladder, an isolated dimer model consistently describes the observations very well. The analysis of the temperature dependent magnetization data leads to a magnetic exchange constant of $J$~=~43~K along the rungs of the ladder and an average value of the $g$-factor of 2.25. From the ESR measurements, we determined the single ion anisotropy to $D$~=~11.5~K. 
The validity of the isolated dimer model is supported by Quantum Monte Carlo calculations, performed  for several ratios of interdimer and intradimer magnetic exchange and taking into account the experimentally determined single ion anisotropy. The results can be understood in terms of the different coordination and superexchange angles of the oxalate ligands along the rungs and legs of the 2-leg spin ladder. 
 
\end{abstract}


\maketitle

\section{Introduction}

The physical properties of one-dimensional spin systems have attracted a lot of attention.
For isotropic magnetic interactions described in the Heisenberg model and  low spin quantum numbers the ground state properties are strongly influenced by quantum fluctuations and often pure quantum ground states are found in a macroscopic system. For the ideal $S$=1/2 antiferromagnetic
Heisenberg chain ($S$=1/2 AFHC) with uniform nearest neighbor
exchange coupling the ground state is a spin singlet with gapless excitations \cite{Takahashi99,Kluemper00}. Introducing an additional interchain magnetic exchange coupling
between adjacent spin chains the so-called n-leg spin-ladders are formed. 
These have been discussed very intensively for the $S$=1/2 case since they exhibit completely
different ground state properties \cite{RiceScience96,DagottoReview}. In particular for even-leg ladders a non-magnetic spin-liquid ground state with a finite
energy gap  is found \cite{DagottoPhyRevB92}.
  
For $S$=1  already for a single antiferromagnetic chain the ground state is a non-magnetic spin-liquid with a finite
energy gap as proposed by Haldane \cite{HaldanePhysRevLett83} and several experimental realizations
are known \cite{EXPHaldanechains}.  
Only very little is known about the ground state and excitations of a S=1 2-leg spin ladder.
From bosonization studies  \cite{SenechalPhysRevB95,AllenPhysRevB00} and quantum Monte-Carlo simulations \cite{TodoPhysRevB01} a continuous crossover from the delocalized Haldane spin gap state to the case of localized antiferromagnetic dimers, which also shows a spin excitation gap, is predicted.   

Experimentally the situation is complicated by the fact that most $S$=1 systems (usually based on high spin Ni(II) transition metal ions) show  a strong axial crystal field anisotropy $D$
which modifies the ground state of the spin system \cite{ChenPhyRevB03}.
To the knowledge of the authors no realizations of an $S$=1 spin ladder have been reported so far.

In this paper we report the synthesis and structure and in particular  a full experimental and
theoretical characterization
of the magnetic properties of a Ni(II) based $S$=1 2-leg spin-ladder material, Na$_2$Ni$_2$(C$_2$O$_4$)$_3$(H$_2$O)$_2$. We performed magnetic susceptibility and high field magnetization studies (up to 55 T) 
to determine the thermodynamic properties. 
In order to study the spectrum of spin states in this system, we have used tunable high-frequency high-field ESR-spectroscopy at frequencies up to 740\,GHz in magnetic fields up to 30\,T. 

Combining the experimental results from low field magnetic susceptibility, high field magnetization  and ESR we are able to determine all relevant parameters of the spin
system. In particular we find 
that the structurally nearly isotropic $S$=1 spin ladder is magnetically very well described
by isolated spin dimers with an antiferromagnetic exchange of $J$ = 43~K on the rungs and a large
axial anisotropy of $D$ =~11.5~K. 
In order to quantify the degree of intrachain versus interchain magnetic exchange in this system  we performed realistic QMC calculations including the experimentally determined 
single ion anisotropy which support the relevance of the isolated dimer model for this system.     
The calculated field dependence of the spin states suggests that
strong magnetic fields allow to tune the ground state from a non-magnetic $S\,=\,0$ to a magnetic $S\,=\,1$ state and at even higher fields to the fully polarized $S\,=\,2$ state. This is confirmed by our magnetization data which provide direct evidence for a magnetic field induced switching of the $S$=0 ground state to an $S$=1 state at $B\,\approx\,30$\,T
and to $S$=2 at $B\,\approx\,60$\,T~.

\section{Results and Discussion}

\subsection{Synthesis and crystal structure}

The compound Na$_2$Ni$_2$(C$_2$O$_4$)$_3$(H$_2$O)$_2$ (SNOX hereafter) is synthesized in
a hydrothermal reaction from solutions containing a very high sodium halide concentration.
Typically Na$_2$C$_2$O$_4$, Ni(C$_2$O$_4$)(H$_2$O)$_2$, NaBr and distilled water are mixed
in a 1:1.9:140:500 molar ratio, heated in a sealed autoclave to 225 $^\circ$C for 5 hours and allowed
to cool slowly to room temperature. The product occurs as green crystals of Na$_2$Ni$_2$(C$_2$O$_4$)$_3$(H$_2$O)$_2$ embedded in a matrix of solid NaBr. 
%While the product is not thermodynamically stable in pure water, 
%decomposition is slow, and so it may be isolated
%by rapidly dissolving the NaBr in a minimum quantity of water, 
%collecting the product by filtration.
Phase homogeneity and purity of SNOX were established by a combination of optical microscopy,
powder X-Ray diffraction and elemental analysis. By slowing down the cooling rate, and reducing 
the concentration of Ni$^{2+}$ and C$_2$O$_4$$^{2-}$ we can grow single crystals up to 2 mg in lower yielding reactions.

Single crystal X-ray structure determination reveals SNOX to crystallize in the monoclinic space group P2$_1$/c (\#14) with a=5.8144(2), b=15.6474(5), c=6.8357(3) \AA, $\beta$=100.56(60) $^{\circ}$ and V=611.37(4) \AA$^3$ at 120~K. Refinement of 118 parameters gave a goodness-of-fit S=1.074, R1=0.244 and wR2=0.506 on all data, with the largest peak (and hole) in the difference Fourier map 0.383 (-0.390) e\AA$^3$.

SNOX is an end member of the series of isostructural compounds with the composition
 Na$_2$T$_2$(C$_2$O$_4$)$_3$(H$_2$O)$_2$ with T~=~Mn(II), Fe(II)\cite{DJPDalton00}, Co(II)\cite{DJPDalton03} and Ni. Although this structure type has been described previously\cite{DJPDalton00}, it is useful to recap certain key features here.
The Ni(II) ion lies on a general position, and experiences a pseudo-octahedral coordination environment. It is coordinated in a \emph{cis} geometry by two chelating and crystallographically 
independent oxalate dianions. The remaining coordination sites are filled by a 
monodentate oxalate oxygen atom and a water molecule. Ni-O bond lengths lie within the range 
2.020(1) to 2.114(1) \AA. We note that opposite pairs of O have similar Ni-O lengths, 
and that the O$\cdots$O separations between opposite pairs, for which the bond vectors are mostly directed along the crystallographic axes, are (d$_a$=4.163(1), d$_b$=4.044(1), d$_c$=4.082(1) \AA~ ~along the a, b and c-axes, respectively, Fig. 1). The overall structure can be regarded as an anionic 
$[Ni_2(C_2O_4)_3]_n^{2n-}$ network with a 1-D ladder like topology (Fig. 1), where the metal ions form the vertices and the bridging oxalate ions form the \textit{bonds}. The two oxalate 
ions bridge nickel ions in quite different modes. One oxalate forms a symmetric
bis-chelating bridging mode (with a crystallographic inversion located at the C-C centroid) and links
pairs of nickel ions (d$_{Ni\cdots Ni(rung)}$ = 5.294(1) \AA) forming the rungs of the ladder.
The second oxalate is unsymmetrically coordinated, chelating to one Ni(II) ion and forming
a monodentate coordination to a second Ni ion, with 1,3-\textit{syn anti} geometry through 
the bridging carboxylate, this mode forms the ladder legs (d$_{Ni\cdots Ni(leg)}$ = 5.810(1) \AA).
These structural linkages are likely to provide the only significant magnetic superexchange pathways.

\begin{figure}[h]
    \includegraphics[width=0.9\columnwidth,clip]{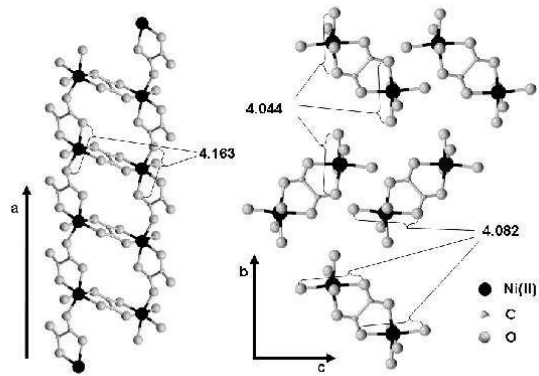}
    \caption{Crystal structure of  Na$_2$Ni$_2$(C$_2$O$_4$)$_3$(H$_2$O)$_2$. For clarity, Na and H atoms are not shown. The numbers indicate the distance (in \AA) between opposing oxygen atoms on the distorted NiO$_6$ octahedra  which are tilted by 17 degree with respect to the a axis.}
    \label{Structure}
\end{figure}

\subsection{Magnetic susceptibility} \label{lowfieldM}

We examined several single crystals %of weight 2.04~mg and  1.79~mg 
in  magnetic susceptibility measurements along the a, b and c axes. The measurements were performed using a Quantum Design Magnetic Properties Measuring System (MPMS) in external fields of $B_{ex}$~=~2~T  and 5~T in the temperature range 2-300~K. In addition we  performed measurements at different angles between the a~axis and the external field by rotating the crystal about an axis perpendicular to the a axis.

\begin{figure}[h]
    \includegraphics[width=0.9\columnwidth,clip]{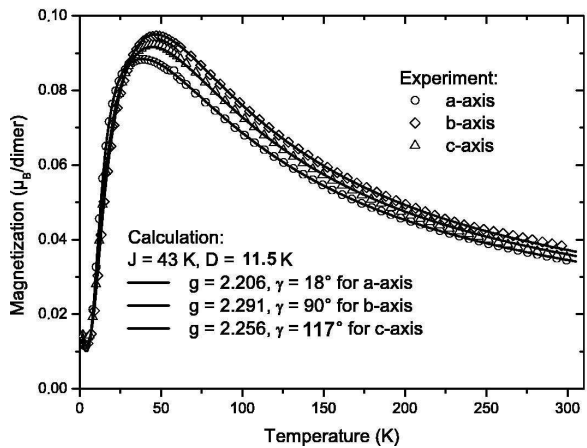}
    \caption{Temperature dependence of the magnetization of a SNOX single crystal (2.04~mg) at an external field of B$_{ex}$~=~2~T along the a~(circles), b~(diamonds) and c~axes (triangles).}
		\label{mvsT}
\end{figure}

Fig. \ref{mvsT} shows the temperature dependence of the  magnetization along the a, b and c~axes in an external field of $B_{ex}$~=~2~T. One can clearly see a pronounced maximum for all directions with a strong downturn below $\approx$ 50 K suggesting a non-magnetic spin-singlet ground state. Notice that the maximum along the a~axes (the ladder direction) is located at a lower temperature than the maxima along the b and c~axes (39~K, 47~K and 45~K for the a, b and c~axes respectively). The difference between the b and c~axes magnetization is interpreted mainly due to a $g$-factor anisotropy.  This behavior in general is  expected as well for an isotropic  two-leg spin ladder as for the two limiting cases, an isolated $S$~=~1 Haldane chain or a system of antiferromagnetically coupled dimers
\cite{HaldanePhysRevLett83,DagottoPhyRevB92,AllenPhysRevB00,SenechalPhysRevB95,TodoPhysRevB01}. 
At very low temperatures a small Curie-tail due to a very low concentration of paramagnetic centers (less than 1 \%) masks the approach to zero. 

Taking into account the two different exchange pathway topologies, we assume a stronger magnetic interaction $J$ along the rungs than the magnetic interaction $K$ along the legs. This leads to an isolated dimer approximation where each dimer consists of two $S$=1 spins on the rungs of the ladder. 
Since all Ni ions on a ladder are crystallographically equivalent they share the same strength and orientation of the single ion anisotropy  $D$.
Including the Zeeman energy in the external magnetic field, the spin Hamiltonian of the system in the dimer approximation is given by
\begin{equation}
    H  = J\, \mvec{S}_1 \mvec{S}_2
           + g \mu_B\,\mvec{B}(\mvec{S}_1 + \mvec{S}_2) 
           + \sum_{i=1,2}\, \mvec{S}_{i}\mmatrix{D}\mvec{S}_i. 
\end{equation}
Here, we assume that the magnetic field is applied along the z-axis, the main axis of the crystal field anisotropy tensor $\mmatrix{D}$. For pure axial anisotropy this tensor is  diagonal, defined as $D_{xx}$=~-~1/3~$D$,  $D_{yy}$=~-~1/3~$D$ and $D_{zz}$=~+~2/3~$D$.  
To calculate the energy levels for applied fields along different directions, we rotate the axis of the axial anisotropy tensor by an angle $\gamma$ about a perpendicular axis. Then the spin Hamiltonian can be expressed as
\begin{equation}
H  =   J\, \mvec{S}_1 \mvec{S}_2
           + g_\gamma \mu_B\,B(S_{1z}+ S_{2z}) 
           +\sum_{i=1,2}  \mvec{S}_i\mmatrix{U}_\gamma^T\mmatrix{D}\mmatrix{U}_\gamma\mvec{S}_i\label{Ham}
\end{equation}
with the rotation matrix $\mmatrix{U}_\gamma$ and an angle dependent g-factor g$_{\gamma}$.

The isolated dimer approximation may be improved by introducing an effective exchange interaction $K$ along the legs treated in mean field approximation in the calculation of the magnetic susceptibility $\chi$ using
\begin{equation}
\chi_{ladder} = \chi_{dimer}/(1+K\chi_{dimer})\label{mfSus}.\\
\end{equation}

To analyse the magnetization data, we developed a fit routine which numerical diagonalizes the Hamiltonian (\ref{Ham}), calculates the free energy $A$ and magnetization $M$ for a 
particular magnetic field orientation $\gamma$ 
%and the deviation from the data 
%(with threefold weighting of the high temperature data T~$\ge$~50~K) 
and varies the parameters in the Hamiltonian to minimize the mean square deviation between the data and this model. We performed a combined fit of $M\,=\,\chi B_{ex}$ for $B_{ex}$~=~2~T along the a, b and c~axes using an isotropic magnetic exchange coupling constant $J$, a fixed absolute value of the axial anisotropy of $|D|$~=~11.5~K (determined by the S=1 zero field splitting measured by ESR as described below). 
The total susceptibility $\chi$ consists of the main
contribution (\ref{mfSus}) and comprises also a temperature independent term $\chi_0\,=\,\chi_{dia}\,+\,\chi_{vv}$ which includes a diamagnetic
contribution $\chi_{dia}$ and a Van-Vleck paramagnetic susceptibility $\chi_{vv}$ of Ni(II), as well as a Curie contribution $C/T$ with a Curie
constant $C$ owing to paramagnetic impurities:
\begin{equation}
\chi = \chi_0 + C/T + \chi_{ladder}. \label{totalchi}
\end{equation}
In addition, we used an orientation dependent g-factor g$_{\gamma}$ and, since the orientation of the principal axis (z axis) of the magnetic anisotropy tensor is not known, the angle $\gamma$ as parameters for the different orientations. 
The fit results for the $B_{ex}$~=~2~T measurements are shown in Fig. \ref{mvsT} as solid lines.

We obtain an intradimer coupling constant of $J$~=~43~K, an interdimer coupling constant of $K$~=~0~K and $g$-values of $g_a$~=~2.206, $g_b$~=~2.291~ and $g_c$~=~2.256~. The angles to the anisotropy axis are $\gamma_a$~=~18$^\circ$, $\gamma_b$~=~90$^\circ$ and $\gamma_c$~=~117$^\circ$, a Curie constant $C$~=~0.0056~$\mu$$_B$K/dimer and a temperature independent 
contribution $\chi_0\,=\,0.0087\,\mu_B$/dimer
has been used. The variations of the $g$-values are consistent with the variation of the distances to the adjacent oxygens, leading to a larger $g$-value with smaller distance. Using  $g_\gamma$$^2$~=~$g_\parallel$$^2$cos$^2$$\gamma$~+~$g_\perp$$^2$cos$^2$$\gamma$~~ \cite{AB} with $g_\perp$~=~$g_b$ and $g_{a}$~=~2.206 we find  $g_\parallel$=~2.197. 
The knowledge of $\gamma_a$, $\gamma_b$ and $\gamma_c$ allows a determination of the orientation of the
anisotropy axis in the crystal. $\gamma_a$~=~18$^\circ$ is explained by an anisotropy axis parallel to the line connecting both opposite oxygen neighbor ions along the a~axis (O$_a$ axis hereafter). This line is tilted with respect to the a axis for all Ni sites (see Fig.~\ref{Structure}, right panel) about an angle of 17$^\circ$  determined by X-ray crystallography. Regarding  $\gamma_b$  we have to consider that with respect to the b-axis the O$_a$ axis is tilted alternating from ladder to ladder, which leads to an average tilting of 90$^\circ$, also in very good agreement with the fitted value. Finally, $\gamma_c$~=~117$^\circ$ can be explained by taking into account the monoclinic structure with $\angle$(a,c)~=~$\beta$~=~100.6$^\circ$. The tilting of 117$^\circ$ is approximately the sum of the angle between a and c~axes and an additional tilting of the anisotropy axis with respect to the a~axis towards the c axis of 16$^\circ$. This is 10$^\circ$ more than what is actually
found for the O$_a$ axis. In view of the expected accuracy for the determination
of the anisotropy axis this is reasonable. In summary, from an analysis of the angle dependence of the temperature dependent magnetization, we can determine the orientation of the principal anisotropy axis to be parallel to the connecting line of the opposite nearest neighbor oxygen ions along the a axis, tilted by 18$^\circ$ with respect to the crystallographic a~axis mainly towards the b axis.

\begin{figure}
    \includegraphics[width=0.9\columnwidth,clip]{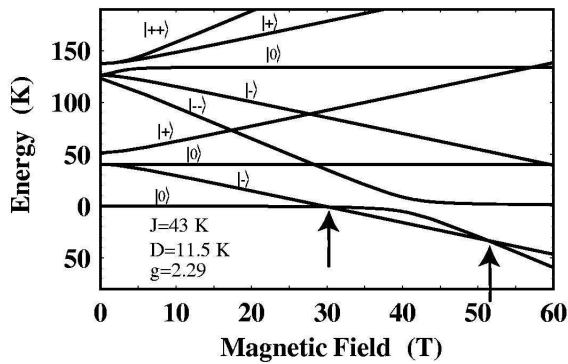}
    \caption{Relative energies of the spin states, calculated for the magnetic field applied  perpendicular to the $z$ axis of the uniaxial
    anisotropy tensor: the triplet $S\,=\,1$ and the quintet $S\,=\,2$ are well separated
    from the $S\,=\,0$ ground state by an activation energy of roughly $|J|\,=\,43$\,K and $3\,|J|$, respectively. Both
    multiplets exhibit a zero field splitting due to crystal field anisotropy. Note the ground state level crossings
    around $30$\, and $53$\, T. (for details see text)}
\label{States}
\end{figure}

The relative energies of the spin states of SNOX calculated in the framework of the Hamiltonian (\ref{Ham}) with the parameters yielding the
best fit of the magnetization are plotted in Fig.~\ref{States} for a magnetic field oriented perpendicular to the axis of the axial anisotropy
tensor. One can see that the spin singlet ground state $S\,=\,0$ is well separated from the $S\,=\,1$ triplet and the $S\,=\,2$ quintet state. The
separation energy between singlet and triplet corresponds to the coupling constant $J$, whereas that between singlet and quintet corresponds to $3J$.
The splitting of the excited $S\,=\,1$ and $S\,=\,2$ states in zero magnetic field is induced by the anisotropy $D$. Both triplet and quintet levels
split with increasing magnetic field due to the Zeeman effect. The role of the zero field and the Zeeman splitting  for the high field magnetic
properties of SNOX will be discussed in detail in Section~\ref{highfieldESR} and \ref{highfieldM}.

Note, that analyzing the data with a negative $D$~=~11.5~K leads to a similar good agreement between the model and the data, resulting in $J$~=~42~K and  $g$-values of $g_a$~=~2.160, 
$g_b$~=~2.256~ and $g_c$~=~2.213~ 
but an inconsistent set of angles $\gamma_a$~=~90$^\circ$, $\gamma_b$~=~25$^\circ$ and $\gamma_c$~=~46$^\circ$ is found. The positive $D$ value is also strongly supported by the high field magnetization data described below.
%This determination of the anisotropy axis is also supported by additional susceptibility %measurements in B$_{ex}$~=~5~T along different angles with respect to the a axis.

An appreciable temperature independent contribution to the static susceptibility $\chi_0$ can not be ascribed to the Van-Vleck paramagnetism
$\chi_{vv}$ alone because typical values of $\chi_{vv}$ for Ni(II) are in the range of 10$^{-4}$~emu/mole \cite{Carlin77} leading to
0.0008~$\mu_B$/dimer at $B_{ex}$\,=\,2\,T. We assign it to a very small concentration of ferromagnetic  Ni metal clusters, which are formed in the
synthesis process as an impurity phase.

\subsection{High Field ESR}
\label{highfieldESR}

ESR measurements in static magnetic fields up to 15\,T have been performed with a Millimeterwave Vector Network Analyzer (MVNA) from AB Millimetre,
Paris. \cite{Dahl98} Experiments in pulsed magnetic fields up to 30\,T have been carried out at the National Laboratory of Pulsed Magnetic Fields,
LNCMP, Toulouse, France, using a Fabry-Perot cavity optically pumped by a CO$_2$ laser
for generation of microwaves. Details of both experimental
set-ups can be found in Ref.~\onlinecite{Golze06}. For the ESR measurements a powder sample of SNOX was prepared as a pressed pellet with a
small amount of a diphenyl-picryl-hydrazyl (DPPH) marker attached for calibration of the magnetic field strength.

\begin{figure}
    \includegraphics[width=0.9\columnwidth,clip]{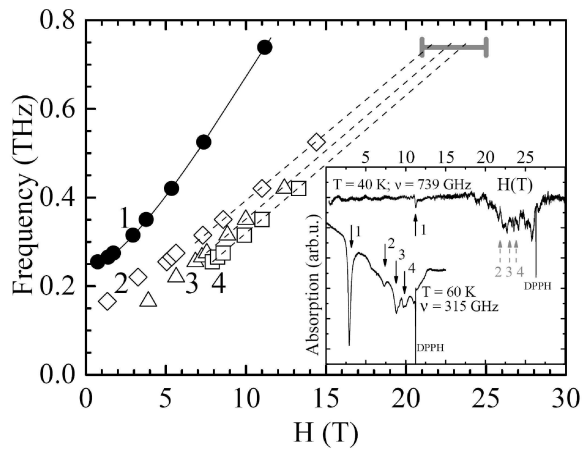}
    \caption{Main panel: Relationship $\nu$ versus $B^{res}_i$ for the ESR modes $i\,=\,1\,\ldots\,4$. Solid line is a theoretical fit of the data.
    Dashed lines is a linear approximation of branches 2, 3 and 4 to high fields. Gray error bar indicates a field region where a broad absorption mode
    is observed in the pulsed field ESR measurement. Inset: ESR spectrum in a static (bottom curve) and pulsed magnetic field ESR experiment
    (top curve). Sharp lines are due to the DPPH field marker. Resonance modes associated with a bulk ESR spectrum of Ni-oxalate are indicated by black arrows and numbered. The anticipated
    positions of modes 2, 3 and 4 at high magnetic fields are marked by gray dashed arrows. (for details see text).}
    \label{ESRmodes}
\end{figure}

\begin{figure}
    \includegraphics[width=0.9\columnwidth,clip]{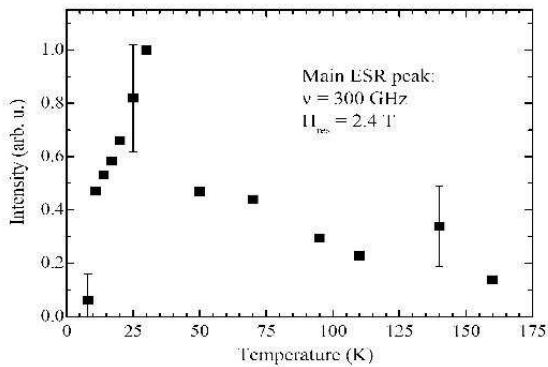}
    \caption{Intensity of the strongest ESR mode 1 as a function of temperature. Note an activating behavior similar to the static magnetization
    $M(T)$ (Fig.~\ref{mvsT}).}
    \label{ESRintensity}
\end{figure}

At a base temperature of 4.2\,K no ESR signals are found which can be associated with the resonance of bulk Ni(II) spins in the sample. However, it
is possible to assign four modes to the bulk ESR spectrum of SNOX which develop in the spectrum with increasing temperature
(Fig.~\ref{ESRmodes}). These modes are numbered from 1 to 4 in order of their appearance in ascending magnetic field and marked by arrows in the
inset of Fig.~\ref{ESRmodes}. The position of the modes in a magnetic field $B^{res}_i$ depends on the excitation frequency $\nu$ as would be
expected for magnetic resonance. Respective resonance branches $\nu$ versus $B^{res}_i$ are plotted in the main panel of Fig.~\ref{ESRmodes}. In the
ESR experiment in static magnetic fields with $\nu\,<\,550$\,GHz mode 1 is the strongest in intensity whereas modes 2, 3 and 4 appear as weak
satellites on the nonlinear background. At a fixed excitation frequency $\nu\,=\,300$\,GHz the intensity $I$ of mode 1 has been studied as a function
of temperature. The result is shown in Fig.~\ref{ESRintensity}. $I(T)$ exhibits an activation temperature dependence similar to that of the static
magnetization $M(T)$ in Fig.~\ref{mvsT}. Because the ESR intensity is determined by the static susceptibility $\chi$ of the spins participating in
the resonance \cite{AB} the common temperature behavior of $I(T)$ and $M(T)\,=\,\chi(T)B$ gives evidence that in the ESR experiment the same spins
are probed which contribute to the static magnetic properties. Mode 1 has a maximum intensity at $T\,\sim\,30\,-\,35$\,K which roughly corresponds to
the energy separation between the singlet ground state and the first $S\,=\,1$ excited state (Fig.~\ref{States}). This allows to unambiguously assign
this mode to a resonance excitation within the $S\,=\,1$ multiplet. Owing to the weakness of modes 2, 3 and 4 they can be detected only in a limited
temperature range $\sim 30\,-\,60$\,K. At smaller temperatures they are not visible due to the strong decrease of their amplitude whereas at higher
temperatures they also broaden. Nevertheless their occurrence in the temperature window where the static magnetization is strongest (Fig.~\ref{mvsT})
allows to assign them to the bulk ESR response of the SNOX sample too.

A remarkable feature of the frequency/field diagram in Fig.~\ref{ESRmodes} is the intercept of the strongest ESR branch~1 with the frequency axis
 at $\nu_0\,\approx\,239$\,GHz. It implies a finite energy gap $\Delta\,=\,\nu_0h/k_B\,=\,11.5$\,K for this resonance excitation. This
zero field gap can be straightforwardly identified with the zero field splitting of the first excited triplet in Fig.~\ref{States}. Numerical
solution of the Hamiltonian (\ref{Ham}) yields $\Delta\,= |D|$. Therefore, from the ESR data one obtains an accurate estimate of the absolute value
of the single ion anisotropy parameter $|D|\,=\,11.5$\,K which is essential for the understanding of the static magnetic properties of SNOX as
discussed in Section~\ref{lowfieldM} and \ref{highfieldM}.

ESR measurements of a powder sample imply the averaging of all possible orientations of the microcrystallites in a powder with respect to the applied
magnetic field $B_{ex}$. As a result orientations perpendicular to the $z$ axis of the uniaxial anisotropy tensor will dominate in the spectrum
whereas other directions may form a broad nonlinear background. Thus for a semiqualitative understanding of the field dependence of the resonance
modes it is sufficient to consider the energy diagram for $B_{ex}\,\perp\,z$ shown in Fig.~\ref{States}. The field dependence of branch~1 can be very
well reproduced (solid line in Fig.~\ref{ESRmodes}) assuming that the resonance transition occurs between the $\left|\,-\,\right>$ and
$\left|\,+\,\right>$ states of the excited triplet (Fig.~\ref{States}). In fact, this usually ''forbidden'' transition \cite{AB} has the strongest
intensity in the spectrum for $\nu\,<\,550$\,GHz. This is due to the strong anisotropy of the crystal field yielding a large zero field gap $\Delta$
and mixing strongly the pure spin states $\left|\,-1\,\right>$, $\left|\,0\,\right>$ and $\left|\,+1\,\right>$ into the $\left|\,-\,\right>$ and
$\left|\,+\,\right>$ states for $B_{ex}\,\perp\,z$ in the field regime $g\mu_BB_{ex}\,<\,\Delta$. Indeed, in a pulsed field ESR experiment the
relative intensity of this mode becomes appreciably smaller (see Fig.~\ref{ESRmodes}, inset). For the excitation frequency of 739\,GHz the resonance
field amounts to $B^{res}_1\,=\,11.2$\,T. For this field strength not the crystal field but the magnetic field direction determines a quantization
axis. Thus the mixed $\left|\,-\,\right>$ and $\left|\,+\,\right>$ states transform into almost pure $\left|\,-1\,\right>$ and $\left|\,+1\,\right>$
states which do not fulfill the ESR selection rule $\Delta S_z\,=\,\pm 1$.

In contrast to mode 1, the exact assignment of modes 2, 3 and 4 is difficult owing to their small intensity. The relationship $\nu$ versus
$B^{res}_1$ has a steep slope because both energy levels, $\left|\,-\,\right>$ and $\left|\,+\,\right>$, depend on magnetic field diverging from each
other with increasing $B_{ex}$. The much smaller slope of branches 2, 3 and 4 suggests that resonance transitions within the neighboring pairs of
states, like e.g. ($\left|\,-\,\right>\,\leftrightarrow\,\left|\,0\,\right>$), ($\left|\,0\,\right>\,\leftrightarrow\,\left|\,+\,\right>$) etc.,  of
the multiplets $S\,=\,1$ and $S\,=\,2$  should be involved. Branch 2 has a much smaller frequency offset as compared to $\Delta$ and can be
associated with the transition $\left|\,--\,\right>\,\leftrightarrow\,\left|\,-\,\right>$ in the quintet $S\,=\,2$. Branches 3 and 4 might be
gapless, which is the case for transitions $\left|\,-\,\right>\,\leftrightarrow\,\left|\,0\,\right>$  in triplet as well as in quintet.
Unfortunately, it was not possible to achieve a better resolution of modes 2, 3, and 4 in the pulsed field measurements. Only a broad absorption mode
is observed in the field region where resonances 2, 3 and 4 are anticipated (Fig.~\ref{ESRmodes}).

\subsection{High field magnetization}
\label{highfieldM}

The high field magnetization measurements were carried out at the  LNCMP, Toulouse, as well. Measurements of the static magnetization have been
performed in pulsed magnetic fields up to 55\,T with a pulse duration of 250\,ms (raising time 40\,ms) by an inductive technique. A system of two
concentric pick-up coils in opposition to each other equipped with additional compensation coils was utilized. The measured signal $\partial
M/\partial B$ was integrated numerically. The sample temperature can be varied between 1.4\,K and 300\,K.

As can be seen from the energy diagram of the spin states of SNOX (Fig.~\ref{States}) the splitting of the states in a magnetic field yields a
level crossing of the ground state with the lowest triplet state $\left|\,-\,1\,\right>$ at a field $B_{C1}$ and a second level crossing of the
$\left|\,-\,1\,\right>$ state with the $\left|\,-\,2\,\right>$ quintet state at field $B_{C2}$ with $B_{C1}\,<\,B_{C2}$. This leads to a step-like
behavior in high field magnetization measurements. If the field is applied along the $z$~axis, the critical field $B_{C1}$ can be estimated using the
equation $B_{C1}\,=\,(J-D/3)/(g\mu_B/k_B$), depending strongly on the sign of $D$. For a field perpendicular to the local anisotropy axis, a negative
$D$ (leading to higher critical fields for $B\,\parallel\,z$) pushes the critical field to lower fields and vice versa for a positive $D$. Since in
powder measurements the perpendicular situation dominates the spectrum of the spin states, the field dependent magnetization, among others, can
determine the sign of the anisotropy $D$.

\begin{figure}[h]
    \includegraphics[width=0.9\columnwidth,clip]{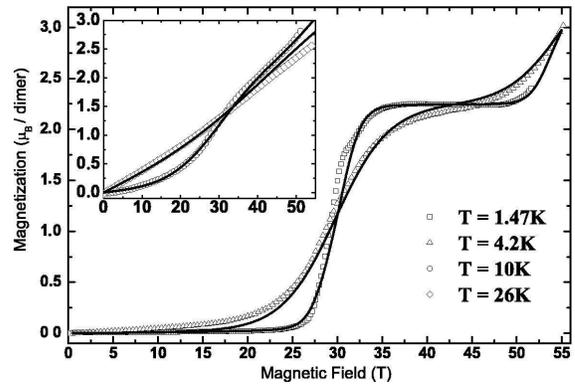}
    \caption{High field magnetization of a SNOX powder sample. The figure shows measurements at 1.47~K and 4.2~K, the inset measurements at 10~K and 26~K. The solid line represents the simulation as described in the text.}
		\label{mvsB}
\end{figure}

\begin{figure}[h]
    \includegraphics[width=0.9\columnwidth,clip]{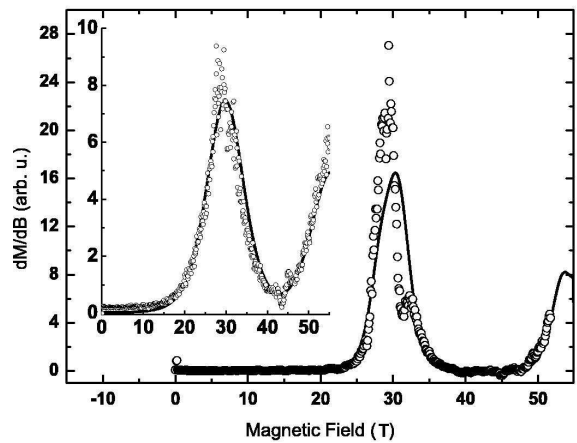}
    \caption{Differential magnetization curves $\partial M/\partial B$ for temperatures T~=~1.47~K and T~=~4.2~K (inset). The open circles are calculated from measured data, the solid lines represent  simulations as described in the text.}
		\label{dMdB}
\end{figure}

The critical field strengths $B_{Ci}$ depend on the magnetic exchange $J$ and an exchange energy of 43~K corresponds to roughly 30~T for $B_{C1}$. Therefore high magnetic fields are needed to observe the magnetization steps. We performed measurements at several temperatures (1.47~K, 4.2~K, 10~K and 26~K) on a powder sample with a mass of~23.2~mg in magnetic fields up to 55~Tesla. The results are shown in Fig.\ref{mvsB}. In  the low temperature experiments the first magnetization step is clearly seen, whereas it smoothes at higher temperatures due to the thermal averaging process. The critical field $B_{C1}$ can be obtained from the derivation of the magnetization dM/dB which is shown in Fig.\ref{dMdB} for $T$~=~1.47~K and $T$~=~4.2~K (inset). At both temperatures a sharp peak is observed at 29~Tesla. Additionally a second increase of the magnetization at much higher fields of about 50~Tesla can be anticipated. This step is also observed strongly broadened in the 10~K measurement. At 26~K a linear field dependence of the magnetization is found.

The solid lines in Fig.\ref{mvsB} and Fig.\ref{dMdB} describe simulations using the dimer model with the calculated g values $g_\parallel$ (see above) and $g_\perp$~=~$g_b$ in a powder average $M$~=~($M_\parallel$+2~$M_\perp$)/3 . To describe the high field magnetization well we need to use a coupling constant of $J$~=~44~K instead of $J$~=~43~K. In particular the differential magnetization is very well reproduced by  the simulation, describing as well the width as the center position of the peak. A small deviation is found in the field range of 30~-~33~T in the $T$~=~1.47~K measurement, which may depend on a little shift of the sample induced by the external field. 

The simulations for a negative $D$~=~-11.5~K did not describe the data reasonably well for any $J$. In this case the first magnetization step is consistent with $J$~=~43~K, whereas the second step is predicted at too low a field. To reproduce the second step well one has to use $J$~=~45~K which results in a too large a 
value of $B_{C1}$.

From Fig.\ref{dMdB}, we can determine the critical field $B_{C1}$ of the powder sample to $B_{C1}$~=~29~T. In our simulations, we get $B_{C1}$~=~27.7~T for $B$ parallel to the magnetic anisotropy axis  and $B_{C1}$~=~30.7~T for B perpendicular to the magnetic anisotropy axis, resulting also in $B_{C10}$~=~29~T in a powder average which is shown in the figure. Notice, that the powder averaging leads to a broadening of the magnetization steps. For the second step, we get $B_{C2}$~=~67.1~T and $B_{C2}$~=~51.3~T for B parallel and perpendicular, respectively.  This difference leads in a powder sample (average over all angles, which we calculated for angles from 0$^\circ$ to 90$^\circ$ in steps of 1$^\circ$ with varying g from $g_\parallel$ to $g_\perp$) at $T$~=~1.47~K to a continuous double-sigmoid like increase of the magnetization in the field range of 50~-~70 Tesla, of which only the first increase is observed in the measurement.

\subsection{Quantum Monte Carlo Calculations}

To perform a quantitative analysis of the ratio
R=J/(K+$|$J$|$) of the intra- to inter-dimer coupling, we
have compared the observed magnetic susceptibility and the
magnetization with Quantum Monte Carlo (QMC) calculations. 
These calculations explicitely include the effects of the
finite
single ion anisotropy $D=11.5$~K. This is different
from
previous QMC studies of spin-1 ladders \cite{TodoPhysRevB01}.
The QMC method we employ is the stochastic series
expansion (SSE) \cite{Sandvik91,Sandvik97}, which directly
samples an expansion of the partition function in powers of
$1/T$ until convergence. Within the SSE the cumulants of
this power series are evaluated using diagonal and
non-diagonal updates within the space of exchange operators
from eqn. (\ref{Ham}). The latter update
is greatly improved by introducing directed loop-updates
\cite{Sandvik02}. We have carried out SSE calculations on
systems with up to $200\times2$ sites and inverse
temperatures $\beta\leq 10$.
The system sizes have been checked to represent the
thermodynamic limit within the temperature range studied.
In all figures depicted the statistical QMC errors are much
smaller than the symbol sizes and have been omitted.

\begin{figure}[h]
    \includegraphics[width=0.9\columnwidth,clip]{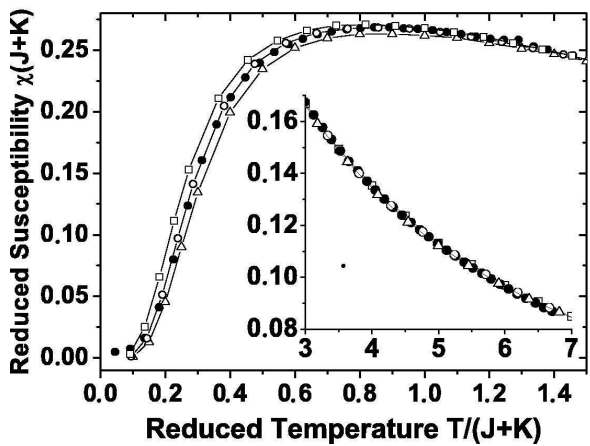}
    \caption{Quantum Monte Carlo simulations for $R$~=~1 (open circles), 0.95 (open triangles) and 0.9 (open squares). We plotted the reduced susceptibility versus reduced temperature. The solid circles show the measured data for $B$ = 2~T along the $a$ axis corrected by a Curie tail and a temperature independent  contribution. The inset shows the high temperature results. 
%The inset shows a comparison between the measured data 
%(crosses) and data measured from Narumi et al on a structural Ni(II) dimer system. 
%We plotted the uniform magnetisation m/m$_{max}$ versus reduced temperature T/J.
}
		\label{QMCsus}
\end{figure}

Our calculations show a significant change of the shape of the susceptibility at low temperatures
as a function of R. In the plot of the reduced susceptibility $\chi_{red}=\chi$~($J$+$K$) versus reduced temperature $T_{red}$~=~$T$/($J$+$K$) the peak increases and narrows with decreasing $R$ down to $R$~=~0.7 and then decreases and broadens with further decreasing $R$ \cite{Simontobepublished}. The calculations have been performed for $D/J$~=~0.27, which corresponds to $D$~=~11.5~K for $J$~=~42.6~K, similar to the fit results described above. The results for $R$~=~1, 0.95 and 0.9 are shown in Fig.\ref{QMCsus} compared with the scaled experimental data. For this scaling we converted the magnetization data for $B$ ~=~2~T parallel to the a~axis to the reduced representation and vary the $g$-factor and the sum of coupling constants $J$+$K$ (which scales $T_{red}$ and $\chi$$_{red}$) to fit the data to the high temperature range (2$\le T_{red}$$\le$8) of the QMC calculations. Notice, that the scaled  experimental dataset appreciably depends on the $g$-factor (in the range of 2.2 to 2.25). Nevertheless, all possible transformations lead to a value of $R$ $\approx$ 0.95 -- 1.00 for total magnetic exchange strenghts between 43~K (R~=~1) and 45~K (R~=~0.95).
The solid circles in Fig.\ref{QMCsus} show the best result of the transformation using a total coupling constant $J$+$K$~=~45~K, a $g$-factor $g$~=~2.23,  a temperature independent 
term of 0.0085~$\mu$$_B$/dimer and a Curie constant of 0.0045~$\mu$$_B$K/dimer. A comparison with the QMC data in the low temperature regime of  $T_{red}$$\le$1.5 gives an  $R$ of 0.95, which results in an intradimer coupling of $J$~=~42.75~K and an interdimer coupling of $K$~=~2.25~K.\\

\begin{figure}[h]
    \includegraphics[width=0.9\columnwidth,clip]{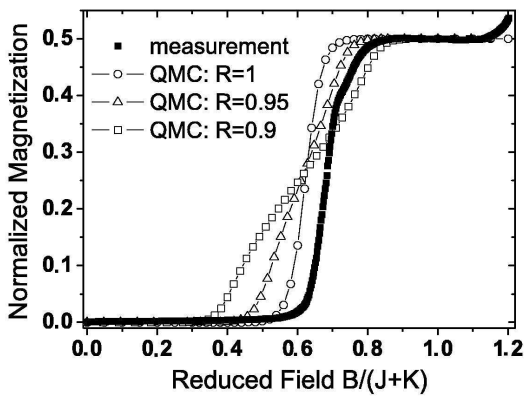}
    \caption{QMC calculations of the field dependent magnetization for different ratios of $R$ with rescpected $D/J$~=~0.27. Open circles: $R$~=~1, open triangles: $R$~=~0.95, open squares:$R$~=~0.9. The calculations were done for a reduced temperature $T/J$~=~0.034, corresponding to $J+K$~=~43~K and  $T$~=~1.47~K. The solid squares are the normalized measurements scaled with  $J+K$~=~43K}
		\label{QMCmag}
\end{figure}

QMC calculations for the high field magnetization of a weakly coupled dimer system  were done for  a reduced temperature $T$/$J$~=~0.034 and a reduced anisotropy $D$/$J$~=~0.27 corresponding to   a sample temperature $T$~=~1.47~K  and $D$~=~11.5~K if $J+K$~=~43~K. The calculations for $R$~=~1, 0.95 and 0.9 are compared with the measured data in Fig.\ref{QMCmag}. The measured data were scaled by setting the magnetization at $B_{ex}$~=~40~T to $M$~=~0.5 and using a value $J+K$~$\approx$~43~K. A variation of $J+K$ mainly shifts the position of the step and does not affect the broadening of the step.
The QMC calculations show a broadening of the magnetization step for decreasing $R$, i.e. increasing magnetic exchange strength along the legs of the ladder. 
The QMC calculations are possible only for $B$ parallel to the anisotropy axis up to now. Therefore, comparing with the powder measurements one has to consider that the magnetization step in the experimental data set is additionally broadened by the presence of the axial anisotropy $D$. Therefore, the step gradient of the 
appropriate QMC calculation has to be sharper than the measurement performed on a powder sample. 
For this reason, the experimental data are in agreement only with the calculation for $R$~=~1, i.e. the uncoupled dimer model.

\section{The anisotropy of the magnetic exchange strength}

An unexpected result of the present study is the very large difference between the strength 
of the magnetic exchange interaction along the rungs ($J$~=~43~K) and legs  ($|K|$~$<$ 2~K) of the spin ladder.
This can be qualitatively understood considering the different coordination and superexchange angles
of the exchange mediating oxalate molecules. 

\begin{figure}
    \includegraphics[width=0.9\columnwidth,clip]{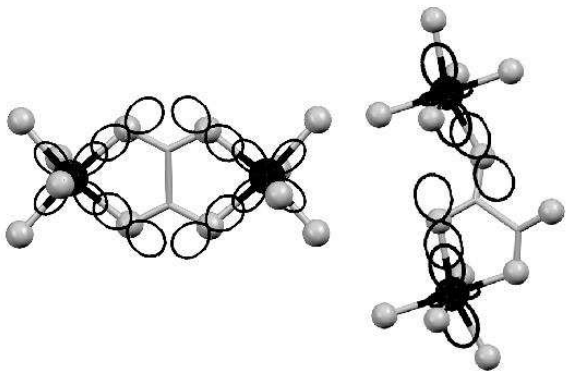} 
    \caption{ Ni-(ox)-Ni coordination and proposed magnetic exchange pathways on the rungs (left panel) and  on the legs of the spin ladder (right panel). The involved Ni 3d and oxygen 2p orbitals are shown in black. }
\label{coordination}
\end{figure}

On the rungs the oxalate forms a $\mu$-1,2,3,4 bridge between two Ni ions. This bridge provides two symmetric superexchange pathways. On each path the Ni 3d x$^2$-y$^2$ orbitals have an enhanced electron probability density extending directly towards the corresponding oxygen ions. There is a direct overlap of the Ni 3d x$^2$-y$^2$ and the O 2p wave functions forming $\sigma$ bonds. The polarized 2p orbitals themselves are strongly overlapping. Therefore the intermediate carbon atom is not involved in the superexchange mechanism   resulting in a strong antiferromagnetic superexchange interaction (see fig. \ref{coordination}). In the literature several $\mu$-1,2,3,4 oxalato-bridged Ni(II) dimer and chain systems are reported  (see e.g. \cite{escuerICA95,romanIChem96} and references therein) with $J$ values in the range of 20 to 42~K. The value of 43 K in SNOX is at the upper limit of the reported range. 

For the  oxalate bridge along the legs of the spin ladder the situation is completely different.
In this case a $\mu$-1,2,3 oxalato bridge is formed. As shown in the right panel of fig. \ref{coordination} only one Ni ion forms two covalent bonds (one with the x$^2$-y$^2$ orbital and one with the 3z$^2$ - r$^2$ orbital) with  the oxalate molecule whereas of the second Ni ion only the 3z$^2$ - r$^2$ orbital forms one covalent bond. The O 2p orbitals involved in the Ni-O bonds are not overlapping with each other and therefore either the carbon atoms or  orthogonal O 2p orbitals are involved  which strongly suppresses the strength of the superexchange mechanism and may even lead to a weak ferromagnetic
exchange. Similar $\mu$-1,2,3 oxalato bridges are reported for Cu(II) systems with magnetic exchange strengths in the range of -0.2 to 0.3~K \cite{nunezICA01,castilloICC01} consistent with the estimate of $|K|$~$<$ 2~K for SNOX.

\section{Conclusion}

In this paper we report the synthesis, crystal structure and magnetic properties of the structurally well defined $S$~=~1 spin ladder compound Na$_2$Ni$_2$(C$_2$O$_4$)$_3$(H$_2$O)$_2$.  We performed magnetic susceptibility and pulsed high field magnetization measurements to examine the thermodynamic properties of the spin system. The magnetic excitations have been examined in high field high frequency ESR. Although the Ni(II) ions form a structural ladder, an isolated dimer model on the rungs describes the observations very well. The analysis of the temperature dependent magnetization data leads to a magnetic exchange constant of $J$~=~43~K along the rungs of the ladder. We deduce the principal axis of the axial anisotropy tensor. From ESR measurements, we determined the single ion anisotropy to $D$~=~11.5~K.
To prove the validity of the isolated dimer model, we compare the single crystal magnetic susceptibility measurements with Quantum Monte Carlo calculations for a $S$=1 spin ladder, which were carried out for several ratios between interdimer and intradimer exchange and  take into account the experimentally determined single ion anisotropy $D$. It results in  a upper limit of 5 \% interdimer exchange coupling. 
This strong anisotropy of the magnetic exchange interaction can be traced back
to the different superexchange pathways on the oxalate molecules along the rungs and
legs of the spin ladder.

\section{Acknowledgments}

We thank A.U.B. Wolter and S. S\"ullow (TU Braunschweig) for help and fruitful discussions in  the
early stage of this work.
Financial support by the Deutsche Forschungsgemeinschaft through SPP~1137 ''Molecular Magnetism'' grant KL 1086/6-1 is gratefully acknowledged. The work of R.K. in
Toulouse was supported by the DFG through grant KL 1824/1-1. Experiments of C.G. and
V.K. in Toulouse were supported by the EuroMagNET consortium of the European Union through contact FP6 R113-CT2004-506239.
D.J.P. is grateful to the EPSRC of the UK for the award of an Advance Research Fellowship (Gr/A00836/02).
We are also indebted to M.E. Light (University of Southampton, UK) for X-ray data collection.

%\begin{thebibliography}

%\end{thebibliography}
\end{document}